\documentclass[conference]{IEEEtran}
\IEEEoverridecommandlockouts
\usepackage{cite}
\usepackage{amsmath,amssymb,amsfonts}
\usepackage{algorithmic}
\usepackage{graphicx}
\usepackage{textcomp}
\usepackage{xcolor}
\usepackage{url}
\usepackage[]{mdframed}
\usepackage{textcase}
\usepackage{caption}
\usepackage{hyperref}
\hypersetup{%
    pdfborder = {0 0 0}
}
\DeclareCaptionFormat{lowercase}{\MakeLowercase{#1#2#3}}
\newcommand{\revision}[1]{{{#1}}}

\def\BibTeX{{\rm B\kern-.05em{\sc i\kern-.025em b}\kern-.08em
    T\kern-.1667em\lower.7ex\hbox{E}\kern-.125emX}}
\begin{document}








\title{Exploring the Role of AI Assistants in Computer Science Education: Methods, Implications, and Instructor Perspectives}


\author{
\IEEEauthorblockN{Tianjia Wang, Daniel Vargas Díaz, Chris Brown, Yan Chen}
\IEEEauthorblockA{
\textit{Department of Computer Science, Virginia Tech, USA} \\
\{wangt7, danielvargasdiaz, dcbrown, ych\}@vt.edu}
}

\maketitle

\begin{abstract}

The use of AI assistants, along with the challenges they present, has sparked significant debate within the community of computer science education. While these tools demonstrate the potential to support students' learning and instructors' teaching, they also raise concerns about enabling unethical uses by students. Previous research has suggested various strategies aimed at addressing these issues. However, they concentrate on introductory programming courses and focus on one specific type of problem. 

The present research evaluated the performance of ChatGPT, a state-of-the-art AI assistant, at solving 187 problems spanning three distinct types that were collected from six undergraduate computer science. The selected courses covered different topics and targeted different program levels. We then explored methods to modify these problems to adapt them to ChatGPT's capabilities to reduce potential misuse by students. Finally, we conducted semi-structured interviews with 11 computer science instructors. The aim was to gather their opinions on our problem modification methods, understand their perspectives on the impact of AI assistants on computer science education, and learn their strategies for adapting their courses to leverage these AI capabilities for educational improvement. The results revealed issues ranging from academic fairness to long-term impact on students' mental models. From our results, we derived design implications and recommended tools to help instructors design and create future course material that could more effectively adapt to AI assistants' capabilities.
\end{abstract}


\begin{figure*}[t]
\centering
\includegraphics[width=1.3\columnwidth]{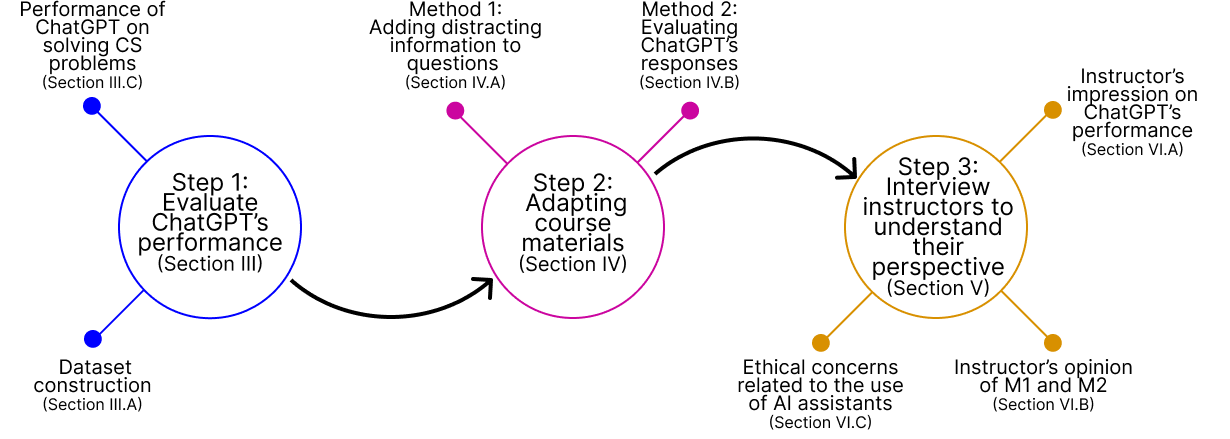}
\caption{An overview of our study workflow where we (Step 1) evaluated ChatGPT’s performance on 187 CS problems, (Step 2) explored two methods to change course materials, and (Step 3) conducted interviews with 11 instructors to understand their perspectives about using AI assistants in CS education.}
\label{fig:main_diagram}
\end{figure*}

\section{Introduction}

Artificial intelligence (AI) is rapidly transforming the way computer science (CS) is taught and learned. Prominent AI assistants, such as ChatGPT~\cite{chatgpt} and GitHub Copilot~\cite{copilot}, provide students with access to advanced problem-solving resources. An increasing number of researchers have shown these AI assistants outperform students when solving complex computing tasks~\cite{jalil2023chatgpt}, and can circumvent plagiarism detection software~\cite{finnie2022robots}. 
These advancements bring about concerns regarding cheating and the integrity of assignments and exams, as students who do not use these assistants may be at a disadvantage when learning new concepts.
Furthermore, AI assistants may generate incorrect answers which could lead students to form incorrect mental models of a concept~\cite{mollick1}. These issues present a challenge for instructors who lack the support to address the impact these AI assistants have on education.

To address these challenges, researchers and organizations have proposed various solutions. For example, studies have investigated strategies to change course materials using prompt engineering~\cite{denny2023conversing}.
Ethan Mollick, a Professor at the University of Pennsylvania, fully embraced AI for his classes by asking students to use AI tools in various ways~\cite{mollick1}. Some organizations have prohibited the use of these assistants~\cite{castillo_2023}, or developed detectors to mitigate their use~\cite{zerogpt,openAIDetector,hugging}. 
However, they have not explored the AI assistants' impact across different topics and program levels of CS courses.
Also, given the rapidly evolving nature and widespread accessibility of AI assistants, these practices are also not feasible in the long run~\cite{chaudhury2022there}. 
Therefore, we ask \textbf{\textit{how can we support instructors to more effectively adapt CS education (e.g., materials, and practices) to the capabilities of AI assistants to prevent students from misuses and improve students' learning experiences}}.

A three-phase study was conducted to answer this question, as outlined in Fig~\ref{fig:main_diagram}. \revision{We utilized a contextual inquiry approach in executing our research~\cite{lazar2017research}.} First, we evaluated ChatGPT's performance in solving problem sets from six undergraduate CS courses, which encompassed a variety of problem types. Second, we explored two problem modification methods based on previous research, including adding distracting information and altering a problem's format and evaluation, to assist instructors in adapting course materials to ChatGPT's capabilities and mitigate potential misuse. Finally, we conducted semi-structured interviews with CS instructors to gauge their understanding of ChatGPT's capabilities, their opinions on our problem modification methods, and their perspectives and concerns regarding the use of AI assistants in CS education.

Our findings show mixed feelings and concerns from the interviewees on issues, such as academic fairness, for the long-term impact of AI assistants. Specifically, we found that 1) while they recognized the potential for students to exploit ChatGPT inappropriately, the majority had not yet made changes to their materials to minimize such misuse because of the lack of effective strategies or tool support; 2) instructors perceived the adapting of course materials to incorporate AI assistants as more feasible for advanced courses than introductory ones; 3) although the interviewees expressed ethical concerns about AI assistant usage, these concerns remained unchanged compared to existing issues; and 4) the false answers generated by ChatGPT could potentially mislead students, causing them to develop incorrect mental models. 

Based on these findings, we discuss two main implications of adapting CS education to AI tools' capabilities, including a) tools and strategies for CS problem adaptation, and b) designing personalized learning experiences using large language models (LLMs).
Our findings contribute to the body of research on the challenges and opportunities that LLM-based AI assistants bring to our society by providing detailed evidence, insights, and analysis from the CS instructors' perspectives, focusing on CS education~\cite{bommasani2021opportunities}.
These ideas work toward a vision of personalized, fair, and adaptive learning experiences for future CS education. This research thus contributes:

\begin{itemize}
    \item Evaluation of ChatGPT's capabilities in solving various types of problems across varying levels and topics of computer science courses, which provides insights into identifying the areas where it may struggle and recognizing its potential use in different aspects of CS education.
    \item Insights of applying two problem modification techniques aimed at assisting instructors in preventing the misuse of ChatGPT by students. The results indicate that the prevailing method of adding distracting information may not be as effective as previously thought.
    \item Design implications and tool recommendations to support instructors in adapting course materials to AI assistants' capabilities. 
\end{itemize}

\section{Background \& Related Work}


ChatGPT is a state-of-the-art large language model capable of engaging in conversations and delivering a variety of assistance to users~\cite{chatgpt}. It exhibits expertise in both natural languages and programming languages. ChatGPT responds to user prompts based on the input it receives and provides answers in real-time. We selected this language model due to its widespread accessibility and cost-free availability.



Recent literature has explored the influence of AI assistants on student learning, such as their use in solving and generating CS problems, and in providing feedback to students~\cite{wermelinger2023using,koutcheme2022towards}. Relevant practical reports, such as the lessons and strategies shared in Ethan Mollick's newsletters, also inform our work, helping shape our strategies and interview questions~\cite{mollick2}.
However, due to the recent public release of these techniques, most studies focus on assessing AI tool performance, not addressing associated concerns. More research is needed on long-term impacts and mitigation strategies.




Several adaption strategies have been proposed to address the issues of academic integrity associated with the use of AI assistants. Companies have explored the development of AI-based cheating detection systems, which aim to identify instances of AI-generated text and code submissions~\cite{zerogpt,openAIDetector,hugging}. However, they often suffer from low  accuracy \cite{pegoraro2023chatgpt}.
Other studies have suggested redesigning course materials to emphasize computational thinking and problem-solving skills, rather than focusing on specific programming tasks that can be easily solved by AI tools \cite{Worell2015,lopez2019experiences,poial2021challenges}.

In addition, researchers have investigated alternative assessment methods that could better measure students' understanding of CS concepts and their ability to apply these concepts in novel situations \cite{Garcia2019,repenning2015scalable}. For example, some have proposed using open-ended projects, collaborative assignments, or interactive programming tasks that require students to demonstrate their understanding of the material in a more authentic and engaging manner \cite{Card2021,Selwyn2019,Zhu2021,guo2015codechella}. Our work contributes to this line of research by examining the performance of ChatGPT on CS problem sets and suggesting modifications to these problem sets to mitigate the impact of AI assistants.

Understanding instructors' perceptions of AI in education is essential for developing effective strategies to address the challenges posed by AI assistants. Prior research has examined instructors' attitudes towards AI-driven tools, such as Intelligent Tutoring Systems, and their potential impact on teaching and learning \cite{woolf1988intelligent,anderson2001role,juarez2013orchestrating,kim2022teacher,al2023acceptance,ji2022conversational}. These studies have highlighted concerns about the potential negative effects of AI tools on students' motivation and learning outcomes, as well as the need for guidance on how to integrate these tools effectively into the curriculum \cite{wang2023preparing,conati2009intelligent,ji2022conversational}. 

However, the new AI-based programming assistants have shown their strong capability on performing these tasks, and there is limited research on instructors' perspectives on their impact on CS education. \revision{Lau's article collected perspectives from instructors on the use of AI assistants in introductory programming courses and explored their short-term and longer-term plans to adapt the courses in response to the AI code tools. \cite{lau2023ban}. Our study investigated the impact of AI assistants on a range of CS courses at various levels and emphasized different implications. We examined two problem modification methods and constructed contextualized examples for semi-structured interviews. This approach facilitated a more concrete understanding of the interviewees' perspectives, concerns, and the potential they perceived for the integration of AI techniques into CS education. Consequently, this allowed us to gather diverse findings and draw unique design implementations, as illustrated in Section Section~\ref{result_section} and Section~\ref{discussion_section}. For instance, we discovered a strong preference among instructors for conducting real-time, in-class activities, but found that existing tools did not adequately support such activities.} Our qualitative findings inform the design implications for future educational materials and instructional strategies, helping to shape a more effective and fair educational landscape.

\begin{table*}[h!]
\centering
\resizebox{1\textwidth}{!}{
\begin{tabular}{l | c c c | c c c | c c c}
\hline
 & \multicolumn{3}{c|}{Multiple Choice} & \multicolumn{3}{c|}{Short Answer} & \multicolumn{3}{c}{Programming} \\
\hline
 & S & PS & NS & S & PS & NS  &S & PS & NS \\
\hline
Introduction to Programming & 75.00\% & 8.33\% & 16.67\%   & 50.00\% & 50.00\% & 0.00\% & 93.33\% & 0\% & 6.67\% \\
Data Structures & 77.77\% & 5.56\% & 16.67\%   & 0\% & 75.00\% & 25.00\% & 75.00\% & 25.00\% & 0\% \\
Computer Organization & 40.00\% & 60.00\% & 0\%   & 72.22\% & 27.78\% & 0.00\% & 28.57\% & 28.57\% & 42.86\% \\
Introduction to Human-Computer Interaction & 48.00\% & 32.00\% & 20.00\%   & 0.00\% & 0.00\% & 0.00\% & 60.00\% & 20.00\% & 20.00\% \\
Computer Security & 100\% & 0.00\% & 0.00\% & 0.00\% & 0.00\% & 0.00\%  & 83.33\% & 16.66\% & 0\% \\
Design and Analysis of Algorithms & 30.77\% & 23.08\% & 46.15\%   & 30.77\% & 23.08\% & 46.15\% & 0.00\% & 50.00\% & 50.00\% \\
\hline
Overall Performance & \multicolumn{3}{c|}{Solvable} & \multicolumn{3}{c|}{Partially Solvable} & \multicolumn{3}{c}{Not Solvable} \\
\hline
 & \multicolumn{3}{c|}{60.85\%} & \multicolumn{3}{c|}{22.75\%} & \multicolumn{3}{c}{16.40\%} \\
\end{tabular}
}
\caption{Percentage of problems in each course that were solvable (S), partially solvable (PS), and not solvable (NS) by ChatGPT.}
\label{tab:results2}
\end{table*}

\section{Data Collection and Performance Evaluation} 
Supplementing prior work, we selected various CS courses across multiple difficulty levels, using collected sample problems from those courses to evaluate ChatGPT's performance. \revision{The dataset is available for download \href{https://docs.google.com/spreadsheets/d/1UmdenyXfzShztOJNjNkVAsV5JyZ-bJA-5EaCv9ZB57Q/edit?usp=sharing}{here}}.

\subsection{Data Selection}
\begin{figure}[!t]
\centering
\includegraphics[width=1.1\columnwidth]{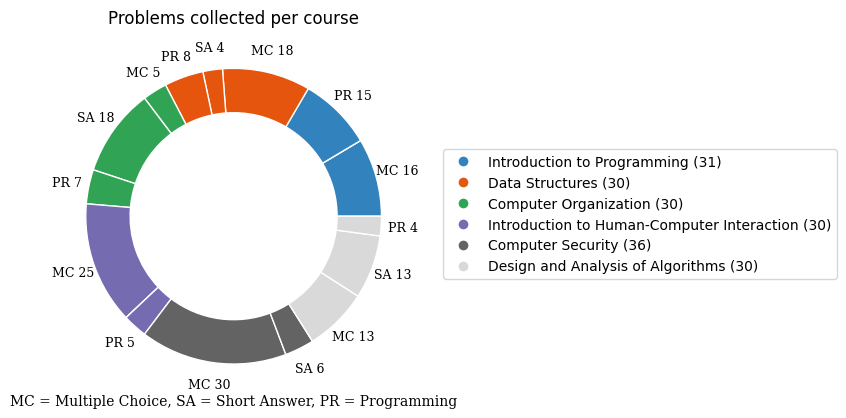}
\caption{Number of multiple-choice, short answer, and programming problems collected from six CS courses}
\label{fig:courses}
\end{figure}

We assessed ChatGPT's performance across six fundamental and advanced CS courses common in degree programs (Figure~\ref{fig:courses}). We collected 30-36 problems per course from public resources like Coursera, Udemy, Udacity, CodeAcademy, and several institutions' course materials, resulting in a total of 187 problems. These problems were chosen for their authenticity and availability of ground truth answers.
The problems, excluding open-ended questions and those with non-text information, spanned three types: \textbf{multiple choice}, \textbf{short answer}, and \textbf{coding}. Multiple choice problems necessitated choosing the appropriate option(s), encompassing Single Answer, Multiple Answer, and True/False categories. Short answer problems required concise text responses, and coding problems necessitated correct code implementation. These categories were chosen for their prevalence across different subjects and to avoid bias in evaluation.

\subsection{Process}
We employed the Jan 30 and Feb 13 versions of ChatGPT, accessible to students for free. Each problem description from the dataset was used as a prompt for ChatGPT under default settings. The generated solutions, influenced by a degree of randomness due to ChatGPT's temperature parameter, were then compared to the correct answers or assessed through manual evaluation or provided test cases. For each problem, we generated three alternative responses, resulting in a total of 561 answers that were manually evaluated by the first two authors. For multiple-choice problems, the generated answer will be compared to the correct option(s) on the associated answer keys. For short-answer problems, the generated answer will be manually evaluated by comparing it with the ground truth answer. For programming problems, the generated answer will be assessed using provided test cases if available in the original problem set, or through manual evaluation by the first two authors. \revision{For each problem, a binary metric was used, meaning that we reported either a full score (1) or a fault (0). The reason behind this is that most of the problems used in the dataset are from online learning platforms, which use a correct/incorrect metric to evaluate answers.}

\revision{We employed ChatGPT to generate three alternative answers for each problem. We defined a problem as ``solvable'' if ChatGPT correctly answered it in all three attempts, each attempt earning a full score of 1 for a total of 3, resulting in an average score of 1. ``Partially solvable'' problems are those for which ChatGPT provided both correct and incorrect answers across the three attempts, yielding an average score between 0.33 and 0.99. A problem was deemed ``not solvable'' if ChatGPT answered it incorrectly in all three attempts, which means that the average score is 0.}

\subsection{Results}
Table~\ref{tab:results2} presents the results of ChatGPT's accuracy in solving CS problem sets across the selected courses. Overall, ChatGPT is able to solve 60.85\% of problems in all three attempts. Our findings demonstrate that ChatGPT can solve various problem types across different levels and topics in CS courses with satisfactory performance. This raises concerns regarding the potential for students to simply copy and paste problems into ChatGPT, obtain solutions, and answer them without truly engaging with the material."

\section{Problem Modification Methods}
Our results from the previous section demonstrate the capabilities of ChatGPT in solving various CS course problems. Due to its proficiency in successfully solving more than 60\% of CS problems in our dataset, which may potentially be misused by students, we explored two methods for helping instructors and TAs modify the problems to prevent inappropriate use of ChatGPT. Method 1 (M1) is to manually adding information or context to mislead or distract the model. The goal of this method is to prevent students from simply copying and pasting problems into ChatGPT to obtain answers. Method 2 (M2) consisted of asking students to validate the answers generated by ChatGPT, which aims to help students to reduce shallow learning and be aware of misinformation.

\subsection{Method 1 (M1): Adding Distracting Information} 

Inspired by prior work \cite{shi2023large}, M1 adds distracting information to a given problem to mislead ChatGPT, rendering the problem unsolvable by ChatGPT to prevent potential misuse. The template of M1 involves manipulating the ``Operation'' and the ``Content'' of the distracting information. The ``Operation'' includes appending, inserting, and editing, while the ``Content'' refers to the definition of a conceptual term related to the problem, related homogeneous information, and supplementary context for the problem.

For example, consider the multiple-choice problem shown in Appendix~\ref{m1_example_original}. ChatGPT can answer it correctly by selecting both B and C with the following response, \textit{``Both B and C are true statements. However, statement A is false. The Edmonds-Karp algorithm is a variation of the Ford-Fulkerson algorithm that uses breadth-first search to choose the augmenting path, which can make it faster in some cases, but not always.''} However, we could apply M1 to add distracting information to make this problem no longer solvable by ChatGPT. As shown in Appendix~\ref{m1_example_midified}, the original problem is presented in black color, while the distracting information is highlighted in blue. After appending the definition of the Ford-Fulkerson algorithm as distracting information to both answers A and C, ChatGPT began to choose B as the only correct answer, and the response now indicates that C is incorrect by saying \textit{``C. The statement is also false. The Ford-Fulkerson algorithm does not have a guaranteed polynomial time complexity, and there exist instances where it can take exponential time to find the maximum flow, even with unit edge capacities.''}

To test the efficacy of this method, we randomly selected 30 problems by taking five problems that are solvable by ChatGPT from each of the six courses in the dataset. We attempted to modify each problem by adding various combinations of ``Operation" and ``Content" to mislead the model. For each problem, we attempted up to 15 distinct combinations, and if none of the combinations were effective in misleading the model, we considered the method to have failed for that problem. Our findings indicate that only 7 out of the 30 problems could be successfully modified to confuse ChatGPT. For those attempts that successfully misled ChatGPT, an average of 8 iterations were needed to explore various combinations of ``Operation" and ``Content."

While performing the evaluation, we noticed some limitations to applying M1. It can only be applied to problems with contexts or textual information. For example, for algorithm problems with only formulas or code completion problems, it would be hard to add distracting information. More identified limitations will be discussed in Section~\ref{perception_on_methods}.



\subsection{Method 2 (M2): Changing The Problem Evaluation} 
Another limitation of ChatGPT is that its responses may not always be accurate due to its inherent probabilistic nature\cite{brown2020language}. If students rely solely on the answers provided by ChatGPT without verifying their accuracy, they may potentially develop flawed mental models \cite{mollick1}. However, this limitation can be turned into an opportunity to enhance the learning experience and prevent the misuse of ChatGPT as M2. To implement this method, instructors can use ChatGPT to generate multiple answers, compare them to the ground truth, and ensure there is at least one incorrect response. After that, instructors can present the original problem along with the ChatGPT's answers to the students, then rephrase the problem to ask students to ``review the problem and answers from ChatGPT, distinguish between correct and incorrect responses, and justify your answer.''

For instance, in the multiple-choice problem shown in Appendix~\ref{m2_example_original}, ChatGPT can be used to generate two distinct answers, with one being correct and one being incorrect. Instructors can then modify the problem format, as shown in Appendix~\ref{m2_example_modified}, by asking students to review the problem with the ChatGPT-generated answers and distinguish between correct and incorrect responses with justifications. 

By using this method, students are encouraged to develop their analytical and reasoning skills, as they must assess the validity of each answer generated by ChatGPT and determine which ones are accurate or erroneous. By having instructors verify the answers and informing students there are incorrect answers to the problem, it could help students avoid learning from the misinformation generated by ChatGPT and developing a flawed mental model. Compared to M1, it also works for problems that do not contain textual or context information. However, a limitation of this method is that it can only be applied to problems for which ChatGPT is capable of generating incorrect solutions. Consequently, simple problems that ChatGPT consistently answers and explains correctly will not be suitable for this approach. More identified limitations will be discussed in Section~\ref{perception_on_methods}.


\section{Interviews}

To gain a deeper understanding of instructors' views on problem modification techniques and their opinions on using AI assistants in CS courses, we conducted surveys and semi-structured interviews with 11 instructors teaching our six selected CS courses. Participants were recruited through a private network within our university and were required to have prior experience as primary instructors for the course. After obtaining their consent, we requested they complete a survey regarding their experience and knowledge of AI assistants. Interviews were scheduled at the participants' convenience, either in-person or online via Zoom, lasting 30-60 minutes. Participants were compensated \$25 for their time and effort. The study was approved by the authors' organization's IRB.

At the interview's outset, we demonstrated ChatGPT's capability to solve problems related to the participants' courses. We then asked if ChatGPT's performance aligned with their impressions and experiences. Next, we provided examples of the two problem modification methods, created based on problems from the course the interviewee had taught, and asked for their thoughts on each. We also discussed their perceptions, ethical concerns, potential applications, and challenges related to ChatGPT in CS courses. The interview sessions were recorded, transcribed using an automated tool, and corrected for inaccuracies. The first two authors coded the interviews and identified significant themes from the transcripts.

\section{Results} \label{result_section}

\subsection{Instructors' Impressions on Performance of AI Assistants}

\subsubsection{\textit{Survey Result}} Prior to the interview, we conducted surveys to gather information about participants' previous experiences with AI assistants. The results show that 63.63\% (7/11) of the participants have prior experience using AI assistants. Additionally, 72.72\% (8/11) of the participants are familiar with the problem-solving capabilities of AI assistants for the classes they currently teach or have taught in the past. From the brief overview of their previous experience utilizing AI assistants the participants provided, we noticed that most participants (8/11) have used AI assistants as a teaching aid or for solving personal tasks. Only 2 out of 11 participants have tried to test the performance of AI assistants in solving problems in the course they taught.

\subsubsection{\textit{Participants were aware of ChatGPT's problem-solving performance in CS course}} To better understand participants' preexisting opinions and general impressions of ChatGPT, we began the interview by demonstrating its performance in solving problem sets in CS courses. We selected 3-5 examples of different problem types from the course taught by participants, including some that could be solved by ChatGPT and others that could not. We then presented these examples to the participants, along with the statistical results in Table~\ref{tab:results2}. After viewing the statistical results and examples provided by us, nearly all (9/11) of the participants reported that ChatGPT's performance in solving CS problems aligned with their previous impressions. One participant (P11) was surprised by ChatGPT's capabilities, which enabled it to solve challenging coding problems that passed all the test cases.

\subsection{Instructors' Perceptions on Proposed Methods} \label{perception_on_methods}

\subsubsection{\textit{Distracting information can be helpful but also confuse students}} Our M1 modifed the problems by using distracting information to mislead or distract ChatGPT. Less than half of the participants (5/11) felt that this method is helpful for making problems harder to solve by ChatGPT, with P2 saying \textit{``it could discourage my students from looking at the answers to these questions''}.

Four participants (P1, P8, P9, P11) raised concerns about the use of this method, whereas two participants (P1, P9) mentioned this method may also confuse students. For example:

\begin{quote}
It is good that confuses ChatGPT, it is bad it probably also distracts students, it's distracting information to them as well as ChatGPT.[...] I understand the goal. But we also are trying to be fair to students and be accurate with them. (P1)
\end{quote}

P8 is concerned that the distracting information could be potentially identified and removed by the students, which let them \textit{``easily evade this technique.''}

In summary, participants' responses suggest that M1 could be an effective way to modify the problem and increase the difficulty of solving it for ChatGPT. However, this approach has limitations, such as potentially distracting students and students can possibly remove the distracting information.

\subsubsection{\textit{Asking students to validate answers can be effective, but would require more effort for grading}} We found that the majority of participants (7/11) considered M2 to be engaging and effective for adapting course materials. Five participants believed it could ``\textit{facilitate critical thinking}'' among students while solving the problems. P2 and P6 believed it could ``\textit{enhance students' understanding of AI in general}''. 

Three participants (P1, P10, P11) teaching introductory level courses, which typically have a large number of enrolled students, pointed out that applying M2 is challenging due to the increased effort required for grading, as one mentioned: 

\begin{quote}
However, I am not sure how this could be scalable [...] as an instructor, that means that I have like 300 generated texts that I have to manually go through which [sic], so I won't use it. (P10)
\end{quote}

The results indicate that instructors in some courses perceive the M2 as an effective method with the potential to enhance students' critical thinking skills and facilitate learning about AI assistants. However, it should be noted that this approach may demand additional effort when assessing student responses.

\subsection{Instructors’ Perceptions and ethical concerns regarding using AI assistants in CS education}

\subsubsection{\textit{Instructors worry about academic integrity problems with AI assistants but are mostly open to using them in class with conditions}} As we demonstrated in Table~\ref{tab:results2}, ChatGPT performs well overall in solving CS problems, and its widespread, free access to all students presents a significant challenge to the education system regarding academic integrity issues. From the results of our interviews, almost all the participants (9/11) acknowledged potential academic integrity concerns posed by AI assistants. However, almost all of them (9/11) were open to students using these tools as supplementary aids rather than completely prohibiting their use in the classroom. As P8 stated, \textit{``I think that [banning ChatGPT] is not a good approach, because it is, it is another wave that we can’t evade. Right? We have to face it.''}

While instructors are receptive to the use of AI assistants, they believe that students should utilize them thoughtfully and within certain constraints. As one participant mentioned:

\begin{quote}
There's some way that we can use it well, to be thoughtful about it and be creative. [...] You have to say, here's where we're okay to use it, and then here's where you can't. (P3)
\end{quote}

P5 felt the idea of using AI assistants in theory courses was ``terrible'' as it will take away the learning benefits gained from ``writing the proofs by doing them from scratch.''
In conclusion, while there are valid concerns about academic integrity, the consensus among participants is that AI assistants like ChatGPT can be valuable educational tools when used thoughtfully and within clearly defined boundaries.

\subsubsection{\textit{Instructors haven't adapted courses for AI assistants, considering more in-class assessments to prevent misuse}}  
One of the main impacts that AI assistants have on the CS education system is that they may necessitate instructors to modify their course materials to minimize potential misuse. This viewpoint aligns with the perspectives shared by other scholars~\cite{cub_chatgpt}. As shown in the previous section, 81.81\% of participants were aware of the potential misuse of AI assistants.  However, we found that only two (P1, P4) participants have modified their course structure and material to adapt to AI assistants. Among the eight participants who have not made any changes, seven are open to making adjustments if they receive proper guidance. As participants may have previously adapted their courses to address plagiarism and cheating, we want to gather insights into methods of adapting their courses to AI assistants. In terms of class dynamics and methods of evaluations, the majority of instructors (7/11) were inclined to administer more in-class activities, quizzes, and exams in the future to prevent the potential misuse of AI assistants. Two (P2, P4) mentioned they are considering implementing oral exams. As P4 mentioned during the interview:

\begin{quote}
I'm thinking like, this is in the future, like, I'm most probably doing everything flipped. Okay. So I'm even going to record my lectures, and then they [students] will come to the class to work on the project, under my supervision and help. (P4)
\end{quote}

In summary, most instructors haven't updated their course materials to adapt to AI assistants. However, they intend to adapt by using more traditional methods, such as conducting in-class activities and in-person evaluation mechanisms.

\subsubsection{\textit{Enforcing fairness was challenging before ChatGPT, and remains (not worsens) relatively unchanged even with the existence of ChatGPT}} We gathered instructors' opinions on the fairness concerns raised by AI assistants during interviews. We observed that while some students may begin using ChatGPT to enhance their grades, others might choose not to rely solely on it to gain a deeper understanding of the material. Additionally, we noted that ChatGPT offers a ``plus'' version, providing subscribers with faster and more accurate access to the GPT-4 model. This could potentially result in inequity regarding educational outcomes and resource distribution, as it might create disparities between students who can afford the upgraded version and those who came from a low social-economic status. Initially, based on this observation, we expected that participants might believe ChatGPT would exacerbate the fairness issue. However, only three (P2, P8, P10) out of 11 participants reported that they think ChatGPT could worsen fairness. 
Contrary to the belief held by some participants that ChatGPT will aggravate the fairness issue, the majority of (8/11) participants stated that ensuring fairness is inherently a difficult endeavor, and ChatGPT does not worsen the problem. As one participant expressed about plagiarism:

\begin{quote}
It's always a concern, even before ChatGPT. [...] The source of the concern is that now instead of stealing the code from you [TAs, instructors, and other sources], they're gonna steal from ChatGPT, so it becomes a different source of the issue. (P3)
\end{quote}

It is intriguing to observe that participants' opinions diverge from our prior impressions. They assert that enforcing fairness was already a challenging task before ChatGPT's emergence, and its presence has not worsened the situation.

\subsubsection{Instructors think false ChatGPT results will potentially lead to flawed mental models} As the previous findings indicate, most instructors are open to students using AI assistants in their courses. However, we identified a potential risk associated with using AI assistants in CS courses, where students could rely solely on answers generated by ChatGPTs without verifying the accuracy of the information provided. Studies have demonstrated that accepting unreliable sources as valid could lead to the development of flawed mental models \cite{chapman_fatal_2001}. As depicted in Table~\ref{tab:results2}, ChatGPT fails to provide correct answers for 38.5\% of the problems. While those generated answers may appear to be credible, they can actually contain misinformation. We discussed this concern with the instructors and found that the majority (9/11) acknowledged the potential for incorrect results generated by ChatGPT to contribute to students' inaccurate mental models, with P3 mentioned:

\begin{quote}
I think also, maybe, as this tool is more widely available in use us having a discussion about it at the beginning of class. [...] It gives the wrong answer sometimes, and you have to be thoughtful about what it tells you, not, maybe not telling them, they should use it completely. (P3)
\end{quote}

Overall, a major concern among the participants is that students may rely on inaccurate ChatGPT-generated answers, potentially leading to the development of flawed mental models. To address this issue, some participants (P3, P11) believe instructors should inform students that answers generated by ChatGPT could be incorrect and contain misinformation.

\subsubsection{\textit{Instructors of introductory level courses are more worried about the potential threats posed by AI assistants}} We observed that instructors teaching introductory level courses express heightened concern regarding the potential risks associated with AI assistants. Interviewees mentioned that there are fundamental concepts essential for designing superior programs and applications. The use of AI assistants could lead students to engage in shallow learning and develop misconceptions, which could impact their performance in advanced courses. One participant shares his concern by saying:

\begin{quote}
I feel worried, because I feel like students may be in their introductory classes, like introductory Python, or Java may use ChatGPT, you [students] could do just to solve any assignment. And then they will not build the true understanding. (P4)
\end{quote}

For high-level courses, instructors feel the problems in their course are less likely to be solved by ChatGPT--and they even support the use of this tool in their classes, with one saying:

\begin{quote}
I mean, it should definitely be used. Like I wouldn't ban anything in my class, especially great resources like this. I just think it's important to understand how it can be applied and what its limitations are. (P6)
\end{quote}

The findings indicate a potential relationship between the tool's utility and the academic level of users. Research has demonstrated notable differences in the mental models of graduate and undergraduate students \cite{coll2003investigation}. Moreover, studies have shown that younger students struggle to identify reliable sources of information \cite{wineburg2016evaluating,leeder2019college,wong2022depends}. 

Overall, instructors are concerned that students with underdeveloped mental models may experience shallow learning when using AI assistants. However, it is plausible that employing AI assistants could lead to improved outcomes for students in advanced courses.


\section{Discussion and Design Implications} \label{discussion_section}
In light of our findings, how can we adapt CS Courses to AI assistants such as ChatGPT to prevent their misuse by students? How can we help learners to make better use of AI Assistants to improve learning experiences? To address these questions, we discuss the design implications in this section.




\subsection{Adapting Curriculum with AI-driven Tools and Strategies}

In contrast to educators from other disciplines and institutes~\cite{castillo_2023}, our CS instructor participants are generally more open to using AI in their educational practices. We surmise that their openness stems from their familiarity with the underlying technology and the potential benefits of enhancing learning experiences and personalizing education. The majority of participants believe it is essential to explore effective methods for utilizing AI assistants while preventing their misuse. As P7 stated, ``\textit{No one is going to say should we not let students use AI if it helps them to learn (...) the argument is always will the AI stop them from learning? Yeah, and if it does stop them from learning, can we ban it somehow?}''

Nonetheless, curriculum modification is a frequent task for educators and teaching staff, and it often necessitates careful planning, acquiring new knowledge, iterative processes, and a significant amount of effort from team members~\cite{mirhosseini2023your}.
Previous research has explored and developed tools to support these endeavors with various objectives, such as preventing students from using answer keys from previous semesters, incorporating feedback from past students, or updating course materials to reflect rapidly evolving techniques, particularly in the field of computer science.
Building on the insights from our study, we found a persistent inclination among educators to revise course content; however, a lack of supportive tools has led to limited advancements in this space. Following are some design implications, grounded in our findings and the existing body of work, that could enhance learning experiences.

\textbf{Enhancing In-class Activities with Support Tools}: Many instructors highlighted their preference for conducting real-time, in-class activities to stimulate and assess student learning, bypassing the need for AI tools. Existing tools such as VizProg, PuzzleMe, and CodeOpticon have demonstrated potential for facilitating effective in-class exercises~\cite{wang2021puzzleme,guo2015codeopticon, zhang2023vizprog}. We propose future work to build upon these foundational tools, enhancing them with capabilities to support diverse, longer format in-class activities, and designing features to deter students from directly copying and pasting problems into other applications, similar to the approach adopted by platforms like Hackerrank~\cite{hackerrank}. 



\textbf{Automate Problem Modification}: Our findings revealed mixed views on the effectiveness of M1 (adding distracting information) and M2 (validating multiple answers) strategies. While some interviewees found them useful, others worried about the labor-intensive implementation, evaluation, and possible student confusion. Also, we found it is easier for instructors to verify answers for multiple choice using correct options in the ground truth and coding problems using test cases, but more effort will be required to manually verify answers for short answer problems due to potential uncertainty in student responses. As both methods required manual verification on short answers either from ChatGPT or students, future systems could provide comparison tools to enhance the visualization of the answers and help instructors efficiently verify correctness across iterations. Furthermore, future work could automate the iterative modify-evaluate process, allowing instructors to focus on decision-making. 


\textbf{Sampling the Generated Answers}: When using M2, we observed homogeneity in AI-generated responses. This could potentially hinder learning outcomes, as ideally, responses presented to the students should exhibit diversity to enhance their understanding of concepts from various angles. Future work could explore prompting techniques to generate diverse answers from multiple perspectives. Drawing on the analogy of stratified sampling strategies, where samples are selected based on characteristics such as style or semantic meaning, future systems could allow instructors to efficiently gauge the similarities of generated responses through a range of metrics, and select the most distinct ones for problem modification.

\subsection{Designing Personalized Learning Experiences with LLMs}


Our research reveals that the accuracy of responses generated by ChatGPT fluctuates depending on the complexity and type of the problem. In line with prior research~\cite{becker2023programming}, our participants expressed concerns about the potential misleading effects of such inconsistencies on learners, and the possibility of creating long-term learning impediments. Although AI tools have inherent limitations, they offer the potential for real-time feedback and personalized learning environments. The challenge, as indicated in previous studies on trustworthy AI~\cite{macneil2023experiences}, is to strike a balance between trust and effective decision-making when interacting with AI assistants. To this end, we propose two directions for the development of AI-assisted learning tools that create a more reliable and personalized learning experience:

\textbf{Facilitating Trust Calibration} ChatGPT currently lacks a mechanism to provide transparency regarding the source or reliability of its responses. Learners are left to their own devices to determine the trustworthiness of the provided answers, a task made more challenging by their incomplete or flawed mental models~\cite{models_Ma}. Trust calibration between users and automated systems is a complex yet essential aspect of creating effective human-computer interactions~\cite{lee2004trust}. Misplaced trust in a misleading response can lead to long-term learning obstacles, particularly for novice learners still building their understanding of the concepts~\cite{coll2003investigation}.

Prior work offers potential solutions to this issue, such as highlighting uncertain tokens in AI system's code completion, emphasizing tokens with the lowest likelihood of being generated by the generative model, or spotlighting tokens that are most likely to be edited by a programmer~\cite{vasconcelos2023generation,khurana2021chatrex}. Building upon these ideas, future work should focus on developing adaptive tools that facilitate effective use of AI-generated outputs, aligning student trust with the veracity of the AI-generated responses. This might involve deploying a conversational agent capable of eliciting a student's mental model of a concept, then generating mental model adaptations based on the uncertainty and accuracy of the information provided by the AI. This way, we can pave the way for a more trustworthy and productive AI-assisted learning experience.

\textbf{Steering AI Tool Usage Among Students} Participants advocated for an approach focused on educating students about the potential short and long-term impacts of AI tools, rather than the laborious and guidance-lacking process of modifying course materials. They believed that by imparting knowledge about the limitations and capabilities of AI tools, students would make more informed and responsible decisions about their usage. Instructors could utilize AI tools to illustrate these impacts through scenario-based examples, thus shaping students' mental models of AI tool usage. For instance, an example might highlight a scenario where relying on an AI to generate the correct prompt consumes more time than understanding the concept and solving the problem independently. This approach encourages students to value knowledge acquisition over blind reliance on AI tools.

Concurrently, in the dawn of AI's integration into education, these intelligent tools can potentially serve as personal tutors, enhancing the learning experience. The longstanding research in Intelligent Tutoring Systems (ITS) underscores the potential of AI to foster engaging and personalized learning experiences~\cite{anderson1986automated}. With recent advancements, future iterations of ITS can move away from reliance on human-crafted feedback and towards a dynamic and adaptive model that scales with students' performance. The future of education could see a shift towards a shared learning experience, where students exchange and learn from each other's interactions with AI. This approach could alleviate the burden of grading and course material creation on instructors, enabling them to focus on areas where their influence could be most impactful.

\subsection{Limitation and Future Work}
While we strove to collect problems across diverse topics and types, the sample size of our dataset is still relatively small, with approximately 30 problems for each course. To enhance the validity of our findings, future work should aim to collect a more extensive dataset encompassing various courses and problem types. \revision{One limitation of the dataset pertains to the metrics used. Since we extracted most of the problems from online platforms, we decided to adopt their same binary evaluation metrics with only correct or incorrect possibilities. Even though ChatGPT can provide partially correct answers, we only reported whether it was correct or incorrect according to the platform's criteria.} Additionally, our participant recruitment method, which relied on personal networks, has limitations, as the interviewees are primarily from our own institutions. In future work, recruitment efforts should target a more diverse geographic range to ensure a broader representation of participants. In terms of evaluating ChatGPT's performance, our current approach does not incorporate advanced prompting techniques. It is possible that the model could achieve better results with them and future studies should explore the use of advanced prompting methods to further assess the model's performance and potential improvements.


\section{Conclusion}
The primary objective of this study is to provide a comprehensive analysis of the impact of AI assistants, such as ChatGPT, on CS education. We examined and evaluated two problem modification methods, innovated from previous research, to prevent potential misuse of ChatGPT by students. Through interviews with instructors, we gauged their understanding of ChatGPT's capabilities, collected their assessment of our problem modification methods, and delved into their concerns regarding the use of AI assistants in CS education. Based on our findings, we suggest design implications to aid instructors in modifying their materials and integrating AI assistants into CS education. Our study contributes to the growing body of research on the challenges and opportunities that AI presents to society, specifically to the ongoing discussion on the use of AI in educational settings, and offers a relevant perspective that can inform and shape future policy, practice, and research in the field of CS education. 
By building on these insights, we aim to advance the understanding of how AI should be used in CS education, and provide guidance for educators seeking to adapt their course to AI assistants' capabilities to mitigate misuse and improve students' learning experiences.

\bibliographystyle{IEEEtran}
\bibliography{refs}
\clearpage
\begin{appendices}
\section{Examples of Problem Modification Methods}
\subsection{Example of Method 1}
\subsubsection{Original problem}
\label{m1_example_original}

\noindent Which of the statements below is true? (Select all that apply) \\
\noindent A.The Edmonds-Karp algorithm is always faster than the Ford-Fulkerson algorithm. \\
B.The sum of the capacities of the edges of a network equals the sum of the capacities of the edges of any residual network. \\
C.The Ford-Fulkerson algorithms runs in polynomial time on graphs with unit edge capacities.

\subsubsection{Modified problem after applying M1}
\label{m1_example_midified}
\noindent Which of the statements below is true? (Select all that apply)\\
\noindent A.The Edmonds-Karp algorithm is always faster than the Ford-Fulkerson algorithm. \textcolor{cyan}{The Ford–Fulkerson method or Ford–Fulkerson algorithm (FFA) is a greedy algorithm that computes the maximum flow in a flow network.}\\
B.The sum of the capacities of the edges of a network equals the sum of the capacities of the edges of any residual network. \\
C.The Ford-Fulkerson algorithms runs in polynomial time on graphs with unit edge capacities. \textcolor{cyan}{The Ford–Fulkerson method or Ford–Fulkerson algorithm (FFA) is a greedy algorithm that computes the maximum flow in a flow network.}\\\\\\\\\\\\\\\\\\\\\\\\\\\\\\\\\\\\\\\\\\\\\\\\\\\\\\\\\\

\subsection{Example of Method 2}

\subsubsection{Original problem}
\label{m2_example_original}

\noindent Given an int8\_t variable named X, if X is divisible by a power of 2, what would be the right most bits in X? Or, what is the bit pattern for X in the right most bits? \\
\noindent A. The rightmost N-1 bits will be 0, where N is $2^N = X$. \\
B. The rightmost N bits will be 0, where N is $2N = X$. \\
C. There is no predicable pattern in the bits. \\
D. The rightmost N+1 bits will be 0, where N is $2^N = X$.

\subsubsection{Modified problem after applying M2}
\label{m2_example_modified}
\noindent Review the ``Problem'' and ``ChatGPT Answers'' below. For the answers generated by ChatGPT, distinguish between correct and incorrect responses, and justify your answer: \\

\noindent Problem:
Given an int8\_t variable named X, if X is divisible by a power of 2, what would be the right most bits in X? Or, what is the bit pattern for X in the right most bits? \\
\noindent A. The rightmost N-1 bits will be 0, where N is $2^N = X$. \\
B. The rightmost N bits will be 0, where N is $2N = X$. \\
C. There is no predicable pattern in the bits. \\
D. The rightmost N+1 bits will be 0, where N is $2^N = X$.\\

\noindent ChatGPT Answers:
\begin{figure}[!h]
\centering
\includegraphics[width=0.9\columnwidth]{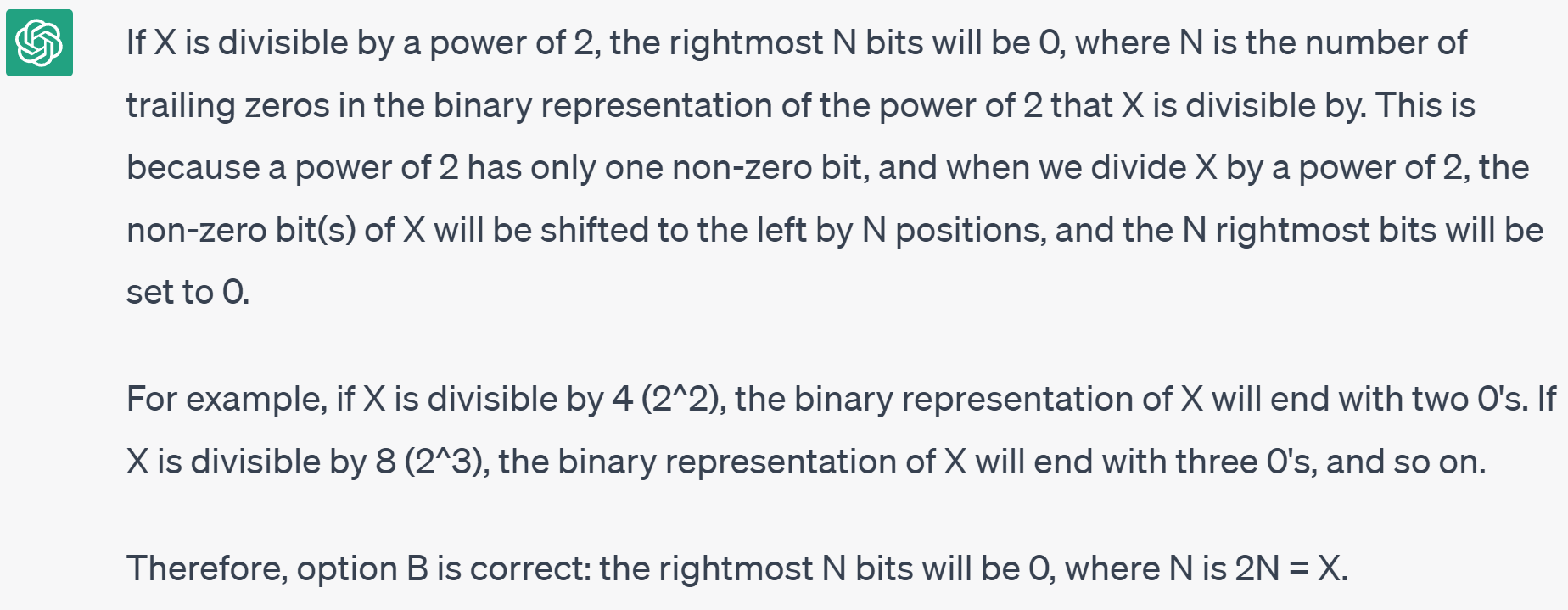}
\label{fig:general_value}
\end{figure}
\\
\begin{figure}[!h]
\centering
\includegraphics[width=0.9\columnwidth]{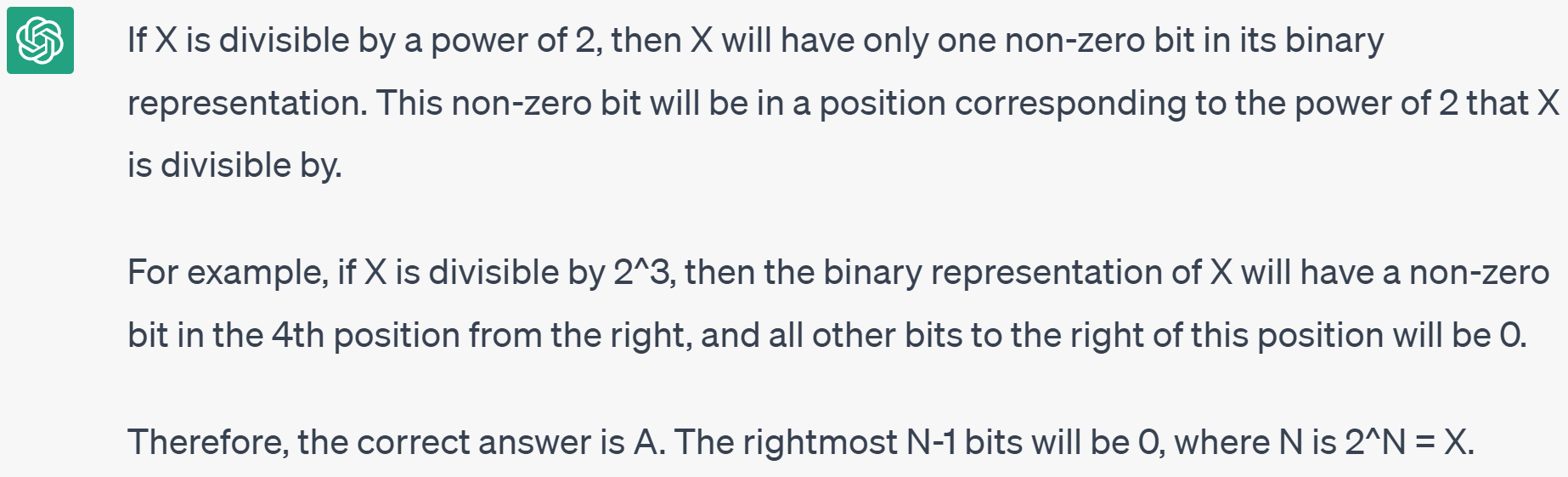}
\label{fig:general_value}
\end{figure}

\end{appendices}

\end{document}